\documentclass[conference]{IEEEtran}
\IEEEoverridecommandlockouts
\usepackage{mathtools,amsmath,amsthm,amssymb,amsfonts}
\usepackage[backend=biber,style=ieee,sortlocale=en-US,url=false,doi=false,date=year,minnames=1,maxnames=5,isbn=false, sortcites, dashed=false]{biblatex}
\usepackage[binary-units=true]{siunitx}
\usepackage{graphicx}
\usepackage{booktabs}
\usepackage{caption}
\usepackage{subcaption}
\usepackage{graphicx}
\usepackage{textcomp}
\usepackage{csquotes}
\usepackage{multirow}
\usepackage{hyperref}
\usepackage{listings,xfp}
\usepackage{anyfontsize}
\usepackage[noend,linesnumbered, ruled, vlined]{algorithm2e}

\newtheorem{example}{Example}

\makeatletter
{\small %
\xdef\f@size@small{\f@size}
\xdef\f@baselineskip@small{\f@baselineskip}
\footnotesize %
\xdef\f@size@footnotesize{\f@size}
\xdef\f@baselineskip@footnotesize{\f@baselineskip}
}
\newcommand{\smalltofootnotesize}{%
  \fontsize
    {\fpeval{(\f@size@small+\f@size@footnotesize)/2}}
    {\fpeval{(\f@baselineskip@small+\f@baselineskip@footnotesize)/2}}%
  \selectfont
}
\makeatother

\renewcommand*{\bibfont}{\smalltofootnotesize}
\AtEveryBibitem{
  \clearlist{publisher}
  \clearname{editor}
}

\title{Orchestrating Multi-Zone Shuttling in \\Trapped-Ion Quantum Computers}
\author{\begin{tabular}[t]{c@{\extracolsep{4em}}c@{\extracolsep{4em}}c@{\extracolsep{4em}}c}
    \large Daniel Schoenberger\textsuperscript{1} & Robert Wille\textsuperscript{1,2,3}
    \end{tabular}\\\\

\small
\textsuperscript{1} Chair for Design Automation, Technical University of Munich, Germany\\
\textsuperscript{2} Munich Quantum Software Company GmbH, Germany\\
\textsuperscript{3} Software Competence Center Hagenberg GmbH, Austria\\
\href{mailto:daniel.schoenberger@tum.de}{daniel.schoenberger@tum.de}, \href{mailto:robert.wille@tum.de}{robert.wille@tum.de}
}

\hypersetup{
	pdftitle={Orchestrating Multi-Zone Shuttling in Trapped-Ion Quantum Computers},
	pdfsubject={},
    pdfauthor={Daniel Schoenberger, Robert Wille}
    }
\begin{document}

\maketitle
\begin{abstract}
Trapped-ion quantum computers are a promising platform, offering high-quality qubits with long coherence times and high-fidelity gate operations. The Quantum Charge Coupled Device (QCCD) architecture provides a scalable blueprint by leveraging the ability to shuttle ions between distinct zones. However, realizing such architectures in practice requires software support to manage ion movement across \mbox{multi-zone} layouts. In this work, we propose a compilation strategy for QCCD architectures with multiple processing zones located outside a \mbox{grid-type} memory zone. Unlike previous approaches that treat processing zones as \mbox{black-boxes}, our method explicitly models their structural constraints, enabling optimized ion movement to and through them. It combines qubit partitioning with \mbox{dependency-aware} gate selection to reduce inter-zone shuttling while enabling simultaneous gate execution.
We implemented the method in an \mbox{open-source} tool and empirically demonstrated its effectiveness across several QCCD layouts, laying a foundation for the compilation of \mbox{multi-zone} trapped-ion systems.
\end{abstract}

\section{Introduction}
\label{sec:introduction}

In recent years, quantum computing has made remarkable progress in scaling and improving available hardware. \mbox{Large-scale} quantum computing promises transformative potential for cryptography~\cite{DBLP:conf/focs/Shor94}, optimization~\cite{DBLP:conf/stoc/Grover96}, and quantum simulations~\cite{PhysRevX.8.011044}. As these research efforts shift from prototypical implementations to more scalable architectures, transitioning beyond the Noisy \mbox{Intermediate-Scale} Quantum (NISQ) era becomes the central goal. Achieving reliable \mbox{large-scale} quantum computation will require further substantial progress in not only hardware but also in developing the necessary software to operate, design, and build new devices. 

Alongside superconducting~\cite{Kjaergaard_2020}, neutral atom~\cite{Henriet2020quantumcomputing, Bluvstein_2023}, and recent candidates such as optical quantum computers~\cite{Slussarenko_2019}, \mbox{trapped-ions} stand out due to long coherence times, \mbox{high-fidelity} gates, and the ability to shuttle ions, which enables flexible qubit connectivity while minimizing additional wiring. In particular, the Quantum Charge Coupled Device architecture has shown promise by leveraging ion movement to employ multiple optimized zones in scalable designs.

QCCD-based devices have already been experimentally demonstrated\mbox{~\cite{Debnath2016, Pino2021, moses2023race}} and experiments on QCCD processors have shown promise for \mbox{fault-tolerant quantum computing}. For instance, \cite{dasilva2024demonstration} has demonstrated how logical error rates can be brought below physical error rates through efficient encoding and error correction on a QCCD device.

Despite these advances, significant challenges remain in orchestrating quantum operations across increasingly complex QCCD systems. Designing scalable QCCD architectures requires efficient and automated \mbox{ion-shuttling} schedules that minimize decoherence and operational overhead, especially when multiple processing zones are incorporated. While prior work has addressed shuttling between single zones and focused on shuttling within the memory zone, the integration and coordination of multiple processing zones introduce additional constraints and orchestration complexity.

In this paper, we address these challenges by presenting a comprehensive compilation strategy tailored to QCCD devices with multiple external processing zones. Unlike approaches that treat these zones as \mbox{black-boxes}, we explicitly model their internal structure. Our compilation framework integrates qubit partitioning with \mbox{dependency-aware} gate selection, reducing the need for shuttling between processing zones and promoting parallel gate execution. We implement our strategy in an \mbox{open-source} compilation tool as part of the \emph{Munich Quantum Toolkit}~(MQT)~\cite{willeMQTHandbookSummary2024} at \url{https://github.com/cda-tum/mqt-ion-shuttler} and demonstrate its efficacy through empirical evaluations across several representative QCCD layouts.

The remainder of this paper is structured as follows: \autoref{sec:background} provides background information; \autoref{sec:general_idea} details the general idea of the compilation framework; \autoref{sec:shuttling} elaborates on the shuttling methods; \autoref{sec:orchestration} presents the orchestration strategy; \autoref{sec:evaluation} describes empirical evaluations; and \autoref{sec:conclusion} concludes the paper.

\begin{figure*}
    \centering
    \subfloat[QCCD device]{\includegraphics[width=.5\linewidth]{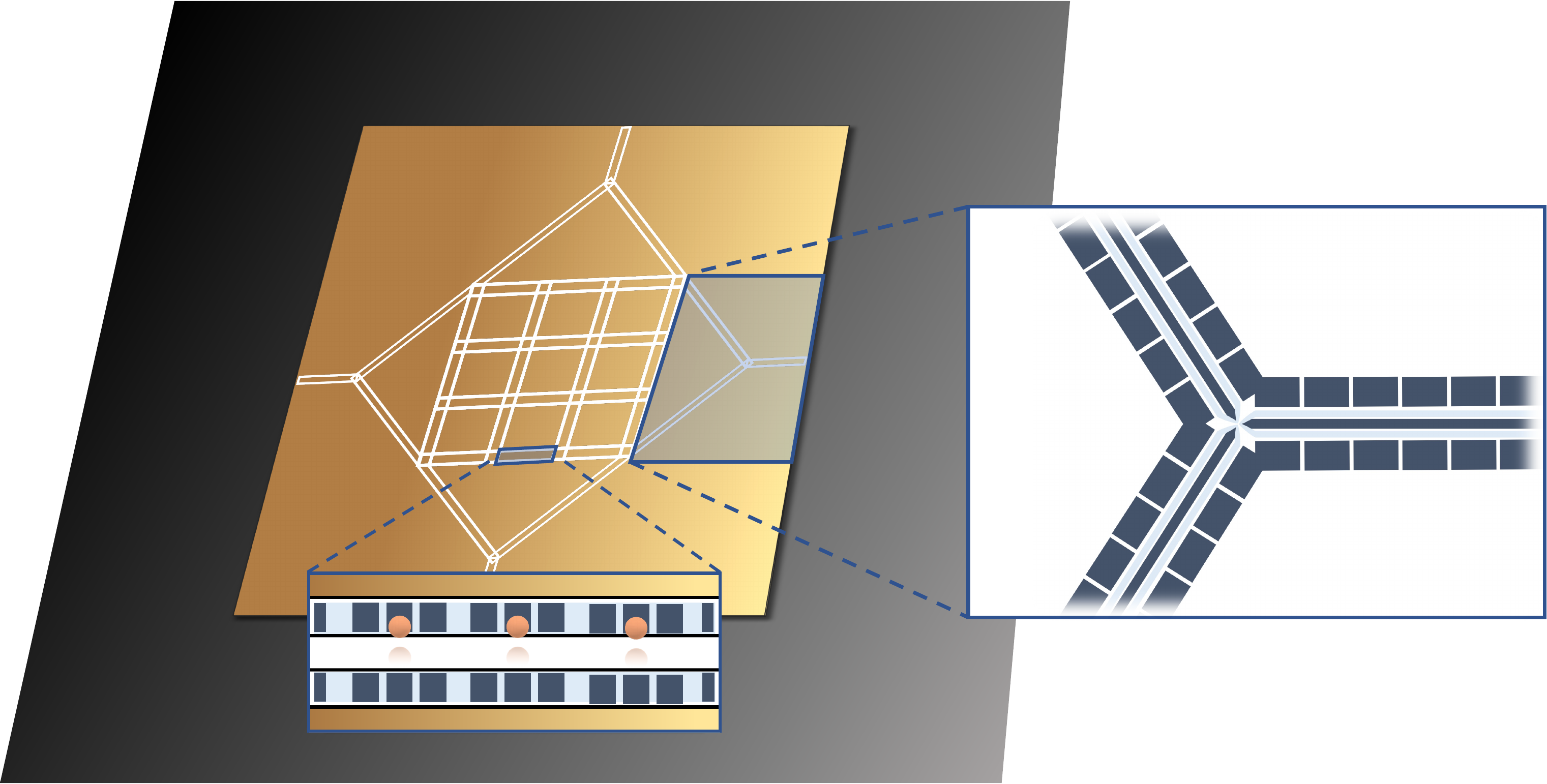}%
    \label{fig:qccd_device}}
    \hfil
    \subfloat[Graph abstraction]{\includegraphics[width=.25\linewidth]{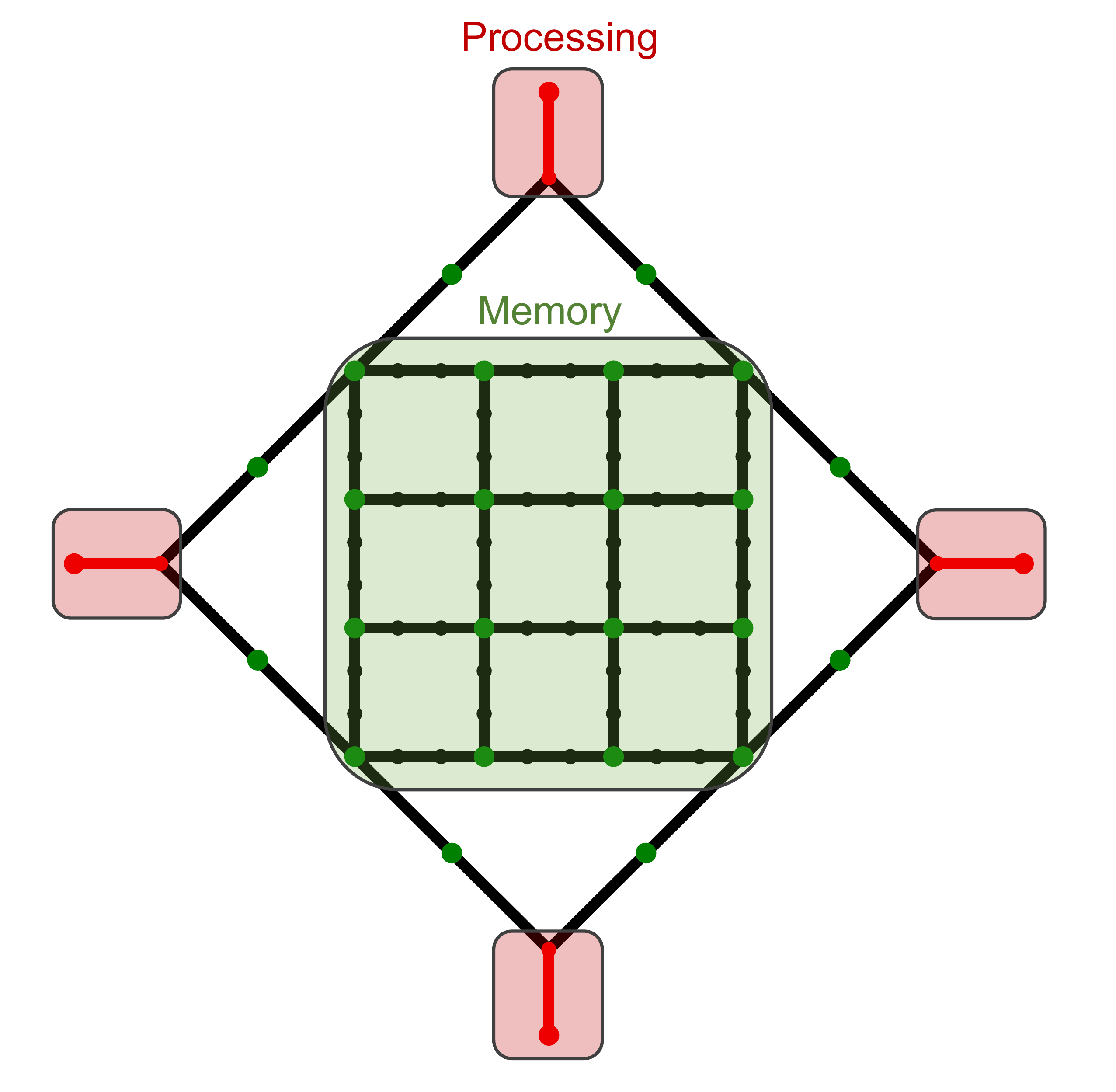}%
    \label{fig:graph}}
    \caption{Illustration of a QCCD device and its corresponding graph abstraction}
    \label{fig:qccd_device_and_graph}
\end{figure*}

\section{Background}
\label{sec:background}

To provide the background for this work, this section reviews the principles of \mbox{trapped-ion} quantum computing and the \mbox{Quantum Charge Coupled} Device architecture, highlighting the aspects most relevant to \mbox{shuttling-based} architectures.

\subsection{Trapped-Ion Quantum Computing}
\label{sec:ion-trap-qc}

\mbox{Trapped-ion} quantum computers utilize individual ions as qubits, confining them via electromagnetic fields~\mbox{\cite{PhysRevLett.113.220501, PhysRevLett.74.4091}}. A common setup used in industrial settings is known as the Paul trap, in which ions are held in a potential generated by a combination of \mbox{radio-frequency} and \mbox{quasi-static} electric fields. By integrating these fields into surface electrode traps, it becomes possible to realize increasingly complex layouts.

A key advantage of \mbox{trapped-ion} technology lies in the ability to physically \emph{shuttle} ions between different locations within the trap. Unlike many other quantum computing platforms that rely on fixed wiring or geometric layouts for qubit connectivity, \mbox{trapped-ion} devices can move qubits as needed, effectively providing \mbox{all-to-all} connectivity. This greatly simplifies connectivity requirements and can reduce the hardware overhead necessary for \mbox{large-scale} quantum processors. However, the act of moving ions must be carefully orchestrated to minimize decoherence and avoid scheduling overhead.

\subsection{Quantum Charge Coupled Device (QCCD) Architecture}
\label{sec:qccd}

The \mbox{QCCD} architecture leverages ion shuttling to build scalable, modular \mbox{trapped-ion} processors~\cite{Kielpinski2002}. Since the ions are able to move around in the system, the QCCD architecture proposes to partition the trap into specialized zones optimized for different stages of quantum computation:
Processing Zones (PZs) for \mbox{high-fidelity} gate operations, Memory Zones (MZs) where ions are stored to protect them from decoherence while idle, Measurement Zones for efficient state readout, and Loading Zones for introducing new ions into the system. 
Under this model, ions shuttle among these zones as required by a quantum algorithm, e.g., starting in the \mbox{memory-zone} for storage, moving to a \mbox{processing-zone} for gate operations, and then returning once they have completed their gate.
To make full use of this potential, future devices may use junctions to connect linear regions and form \mbox{two-dimensional} (2D) architectures.
Due to its concept, a dedicated MZ would not only be shielded from noise but also from control elements such as laser pulses, which limits its operational ability. Consequently, ions within a 2D MZ usually cannot simply swap positions and must rather be rearranged.

\begin{example}
    As a concept of a future 2D QCCD device, see the architecture illustrated in \autoref{fig:qccd_device}. Below the chip, a linear region is highlighted, capable of holding up to three ions (orange). The electrical control elements confining the ions are shown in blue. On the right, the corresponding control elements forming a processing zone (PZ) are depicted, connected via a Y-junction.
\end{example}

Effectively managing these constraints to minimize ion motion is crucial for reducing overall runtime and mitigating the impact of decoherence. Recent device prototypes demonstrate the use of \mbox{multi-zone} capabilities ~\cite{Pino2021,moses2023race}, yet scaling up to large qubit numbers increases the complexity of moving between zones and efficiently exploiting their potential. 

The compilation strategy proposed in this paper addresses exactly these challenges by orchestrating ion shuttling across \mbox{multi-zone} QCCD systems. In particular, we focus on QCCD architectures comprising an MZ structured as a \mbox{two-dimensional} grid and multiple external PZs modeled as linear trap regions. In the following, we consider the shuttling between the MZ and PZs since these are the zones we expect to interact the most while executing a quantum circuit.

\section{General Idea}
\label{sec:general_idea}

\begin{figure*}
    \centering
    \subfloat[Shuttling in the MZ grid]{\includegraphics[width=.27\linewidth]{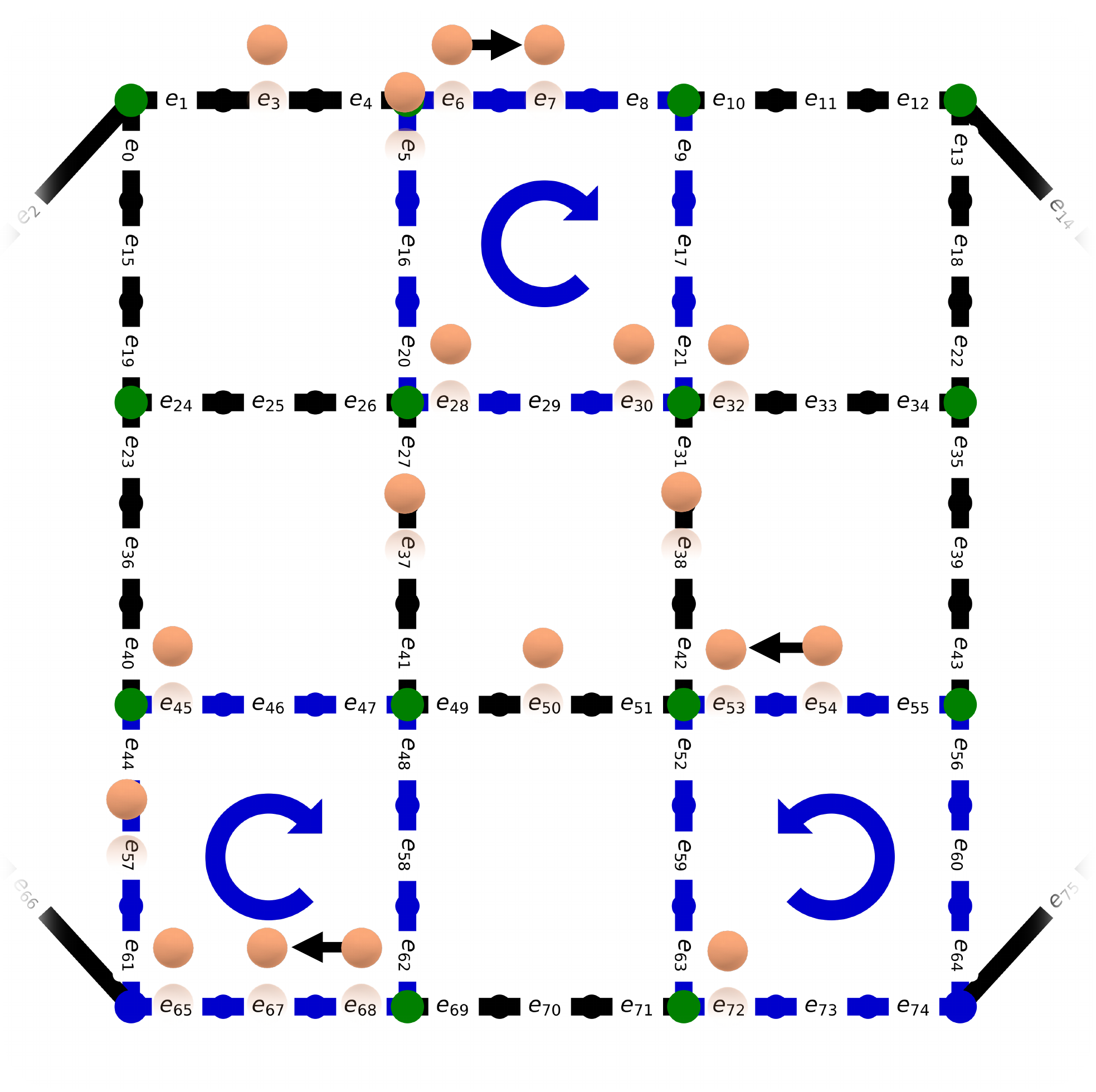}
    \label{fig:grid_shuttling}}
    \hfil
    \subfloat[Shuttling through linear PZs]{\includegraphics[width=.23\linewidth]{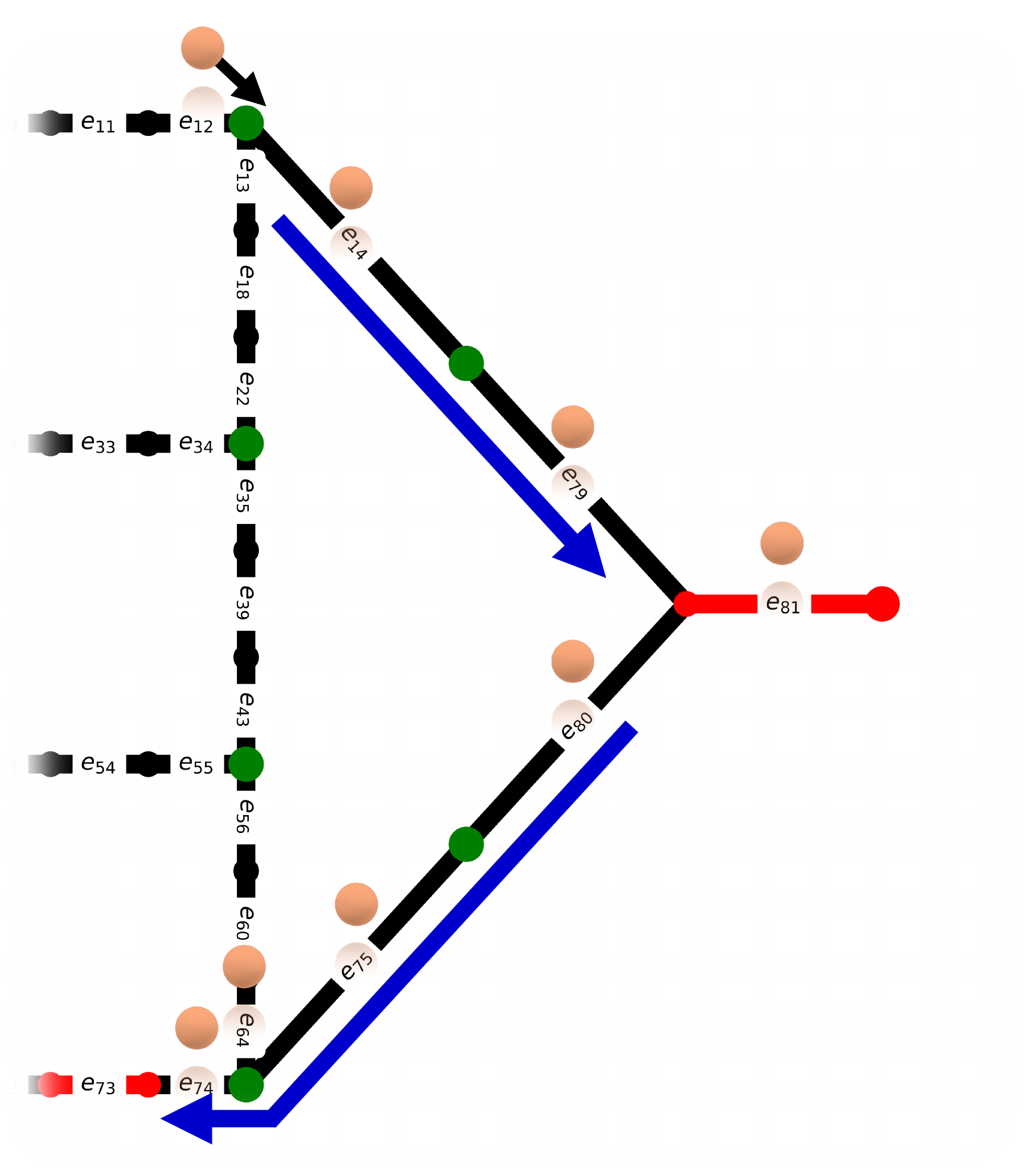}
    \label{fig:path_shuttling}}
    \caption{Comparison of cycle-based shuttling within the MZ and path-based shuttling through a linear PZ.}
    \label{fig:shuttling}
\end{figure*}

The promise of scalable quantum computing using QCCD architectures relies on the ability to efficiently manage ion movement across increasingly complex device layouts. As architectures incorporate multiple specialized zones, the compilation challenge shifts from merely scheduling gates to orchestrating between computation, shuttling, and resource allocation across distributed areas of the device. Efficiently executing a quantum circuit on a QCCD device requires not only efficient shuttling within zones but also coordinated movement between multiple zones, all while maximizing parallel gate execution whenever circuit dependencies permit.

Previous work has mainly focused on optimized movement within smaller systems and specific trap geometries\mbox{~\cite{9138945, Keszcze2021ExactPD, muzzle, Schmale_2022, Kreppel_2023, Drive-Through-Architecture, schoenberger2023using}}. Addressing larger architectures,~\cite{schoenberger2024shuttling} introduced efficient shuttling compilation within a \mbox{grid-type} MZ. However, that work was limited by its connection to only a single PZ for processing quantum gates. Moreover, shuttling within the PZ was treated as a \mbox{black-box} operation, neglecting the internal structure and specific constraints of moving ions into and out of it. 
Managing the simultaneous operation of multiple distinct PZs, each with its own internal shuttling and operational constraints, introduces significant orchestration complexity.
Our approach addresses these challenges with a compilation framework tailored to QCCD architectures featuring a \mbox{grid-type} MZ and multiple PZs. Instead of treating PZs as \mbox{black-boxes}, we explicitly model them as linear trap regions connected at the memory grid boundaries

This framework integrates three core components:

\begin{enumerate}
    \item \textit{Memory Zone Shuttling:} We retain efficient \mbox{cycle-based} shuttling schemes for ion movement within the MZ~\cite{schoenberger2024shuttling}.
    \item \textit{Shuttling through Processing Zones:} For PZ-related movement through linear regions, we introduce and model \mbox{path-based} shuttling, as detailed in Section~\ref{sec:shuttling}.
    \item \textit{Orchestration Strategy:} Crucially, we implement a novel compilation strategy (Section~\ref{sec:orchestration}) to coordinate all \mbox{inter-zone} shuttling using qubit partitioning and gate commutativity to select favorable gates for each PZ.
\end{enumerate}

By combining explicit modeling of multiple PZs with an orchestration layer that manages dependencies and enables parallelism, our approach offers a holistic solution for compiling quantum computations onto realistic QCCD devices and lays the groundwork for efficiently utilizing \mbox{next-generation} \mbox{trapped-ion} processors.

\section{Shuttling}
\label{sec:shuttling}

Efficient shuttling remains the key task in compiling QCCD devices. This section introduces the core shuttling techniques used in this work, on which the next section builds to orchestrate \mbox{multi-zone} shuttling.

\subsection{Shuttling in the Memory Zone Grid}
\label{subsec:mz_shuttling}

Movement within the main MZ, structured as a grid connected by X-junctions, utilizes the cycle-based shuttling approach previously developed in~\cite{schoenberger2024shuttling}.
We briefly summarize the key concepts here.

The MZ architecture is represented as an undirected graph $G=(V, E)$, where the set of edges $E = \{e_0, \dots, e_k\}$ correspond to linear trap segments capable of holding ions, and nodes $V$ represent junctions (major nodes) or intermediate points within linear regions (minor nodes).

\begin{example}
    A corresponding graph representation of a \mbox{two-dimensional} QCCD device is given in \autoref{fig:graph}. Since the device can hold three ions in each linear region, the graph consists of three edges in between junctions.
\end{example}

Moving multiple ions simultaneously along their shortest paths toward a PZ often leads to conflicts, as ions cannot directly swap positions within the MZ. To address this, the \mbox{cycle-based} approach leverages the grid topology. Conflicts are avoided by constructing closed loops (cycles) from graph edges that include both the shortest paths of target ions and any blocking ions. In one time step, all ions on a cycle are rotated forward by one position along the cycle’s direction. This enables target ions to advance while simultaneously moving blockers aside without complex backtracking. Grid architectures naturally offer many rectangular loops, supporting simple cycle construction. Multiple cycles can be executed in parallel per time step, provided they do not overlap, i.e., share a junction or edge. The selection of cycles to execute in case of conflicts is determined by the priority queue generated by the orchestration layer. The concept of the priority queue is explained in \autoref{sec:orchestration}. For further details on the \mbox{cycle-based} algorithm, the interested reader is referred to~\cite{schoenberger2024shuttling}.

\subsection{Shuttling through Linear Processing Zones}
\label{subsec:pz_shuttling}

While cycles work well within the MZ grid, movement into, within, and out of PZs requires a different approach. Our architecture retains the concept of a shielded MZ, connected to PZs via dedicated paths outside the grid. Crucially, instead of treating PZs as \mbox{black-boxes}, we explicitly model them as linear trap regions. To realize the connection and enable directed movement, we utilize \mbox{Y-junctions} as the interface between the MZ and the linear PZs. While more complex PZ interface designs are conceivable, this topology represents a minimal yet functional model incorporating \mbox{one-way} entry and exit paths and an explicit linear processing region. By introducing additional junction nodes into the entry and exit paths, their length is now also considered in the graph representation.

\begin{figure*}[h]
    \centering
    \subfloat[Circuit]{\includegraphics[trim=0cm -10em 0 0, clip, width=.22\linewidth]{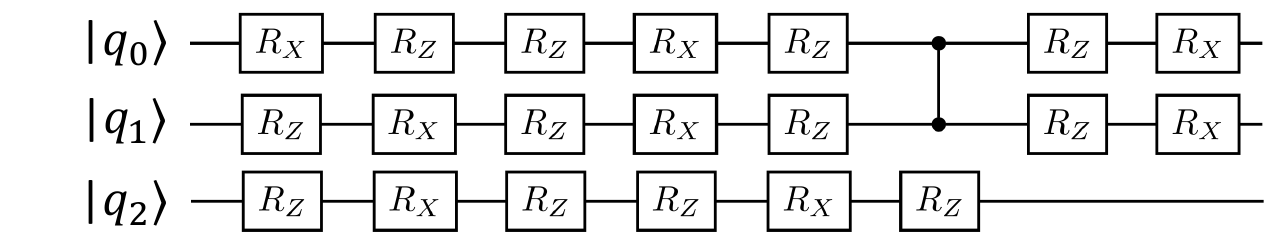}%
    \label{fig:circuit}}
    \hfil
    \subfloat[Dependency graph]{\includegraphics[width=.25\linewidth]{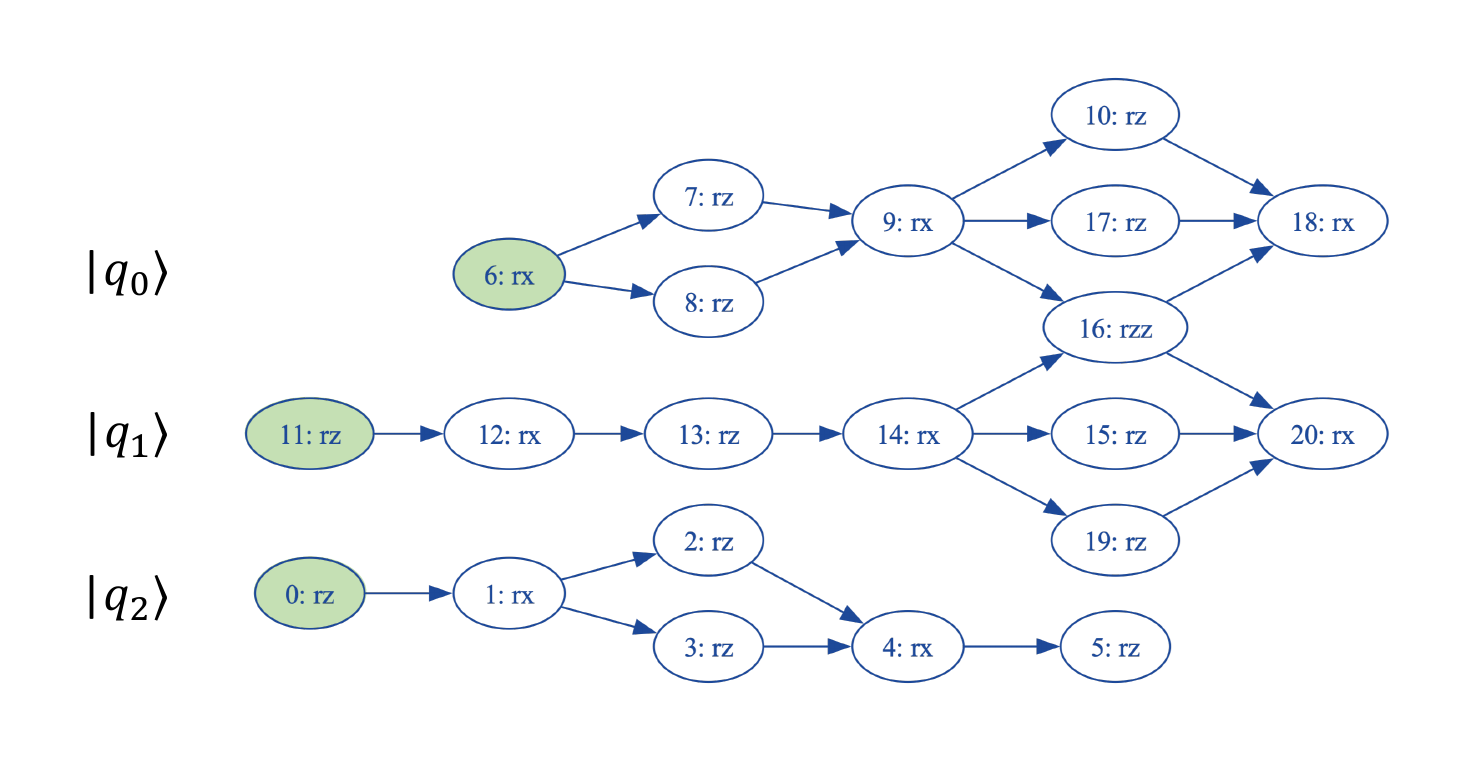}%
    \label{fig:dag}}
    \hfil
    \subfloat[Shuttling schedule]{\includegraphics[width=.2\linewidth]{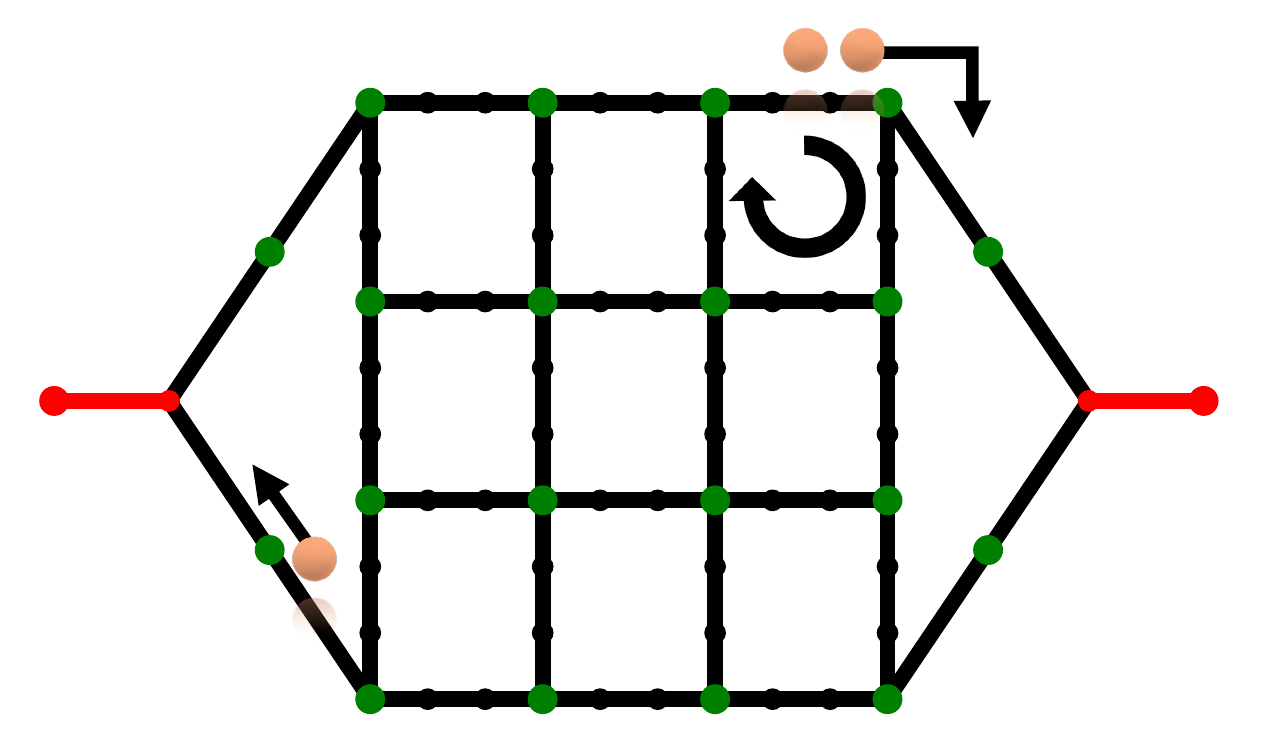}%
    \label{fig:shuttling_schedule}}
    \caption{Compilation steps}
    
    \label{fig:circle}
    \vspace{-1em}
\end{figure*}

To schedule the shuttling through these linear zones, we introduce a \mbox{path-based} approach, that can be seamlessly incorporated into the scheduling of the \mbox{cycle-based} approach. 

\textbf{Movement Towards and Into the Processing Zone:}
When the orchestration strategy (Section~\ref{sec:orchestration}) targets an ion for a specific PZ, and its calculated next move within the MZ would place it onto the linear entry path leading to that PZ, the shuttling mechanism switches from cycle to path generation.
\begin{enumerate}
    \item A directed path is selected from the ion’s current edge along the linear entry path to the target PZ.
    \item In one time step, all ions on this path (including the target ion) are shifted one edge forward.
    \item  Ions can be pushed directly into the PZ edge, provided the PZ is not at its capacity limit.
\end{enumerate}

\textbf{Movement Out of the PZ:}
Unlike in \cite{schoenberger2024shuttling}, multiple PZs may now compete for access to the MZ grid. These exit paths must therefore be integrated into the orchestration scheme (\autoref{sec:orchestration}), which then decides which of the paths are scheduled.

\begin{enumerate} 
    \item To exit a PZ, a suitable unoccupied edge within the MZ is first identified as a target. 
    \item A Breadth-First Search (BFS) is initiated from the MZ junction connected to the PZ's linear exit path to locate the nearest free edge. 
    \item Once found, a directed path is constructed from the ion's current edge through the MZ to the target. 
    \item As with entry paths, all ions along this exit path are shifted one edge forward. 
\end{enumerate}

This ensures ions exiting a PZ reach a clear destination in the MZ. Depending on MZ congestion, locating a free edge and completing the move may take multiple time steps.

\begin{example} 
    Consider the configuration in \autoref{fig:shuttling}. In \autoref{fig:grid_shuttling}, three ions attempt to move but are blocked in the MZ, prompting the construction of three cycles to clear the way. In \autoref{fig:path_shuttling}, an ion is trying to enter the entry to the PZ. To make room for the ion, a path is constructed to the PZ edge. To move ions out of the PZ edge, the path leading from the PZ edge to the free edge $e_{73}$ in the MZ is used.
\end{example}

\textbf{Constraints:}
A key constraint for both entry and exit paths of a PZ is that movement may be temporarily blocked. During quantum gate execution, access to the corresponding PZ is restricted for the duration of the gate. The orchestration layer must account for these timings when scheduling shuttling. Conflicts can also arise if multiple PZs attempt to use overlapping paths, e.g., exiting into the same MZ region. As with cycles, such conflicts are resolved by the orchestration strategy based on priority (see Section~\ref{sec:orchestration}). Scheduling must also respect PZ capacity limits. If a PZ is at maximum capacity, an ion attempting to enter via the entry path will be blocked unless an ion simultaneously exits the PZ.

\section{Orchestrating Multi-Zone Shuttling}
\label{sec:orchestration}

Efficient use of a QCCD architecture with multiple PZs requires a compilation strategy that goes beyond simple gate scheduling. It must orchestrate ion movement between the shared MZ and multiple PZs, respecting circuit dependencies and architectural constraints while maximizing parallel gate execution.
The proposed approach achieves this by integrating two key components: \textbf{(A)} strategic qubit partitioning to minimize shuttling between PZs, and \textbf{(B)} \mbox{dependency-aware} gate selection using a Directed Acyclic Graph (DAG) for each PZ. This combination allows the compiler to make informed decisions that balance costly \mbox{inter-zone} shuttling with the goal of maximizing parallel processing.

\subsection{Qubit Partitioning}
\label{subsec:partitioning}

The first step aims to reduce the overhead from moving ions between PZs. Since PZ connections lie at the boundaries of the memory grid, shuttling between them is typically more \mbox{time-consuming} than waiting for access to an occupied PZ. To minimize such movement, we begin by partitioning the circuit’s qubits across available PZs.
As only one gate per ion can occur per timestep and \mbox{single‐qubit} gates do not introduce dependencies on other qubits, we focus on efficiently scheduling \mbox{two-qubit} gates. Specifically, we aim to group frequently interacting qubits into the same PZ. This is achieved by constructing an interaction graph, where nodes represent qubits and weighted edges capture the number of \mbox{two-qubit} interactions. A repeating bisection strategy based on the \mbox{Kernighan–Lin (KL)} algorithm~\cite{KL} is then applied: starting with all qubits in one set, we iteratively bisect the largest partition until reaching the desired number of PZs or until further splitting is not possible. The KL algorithm heuristically minimizes the cut size of the interaction graph, reducing the number of \mbox{inter-zone} edges and thereby shuttling. Simultaneously, repeating bisection ensures balanced partition sizes across PZs.
\vspace{-.1em}
\begin{example}
\label{ex:partitioning}
    Consider the quantum circuit in \autoref{fig:circuit}. In the case of two available PZs, the proposed strategy focuses on the single \mbox{two-qubit} gate acting on qubits $q_0$ and $q_1$. The two ions representing these qubits are mapped to the first PZ, while $q_2$ will be scheduled to the second PZ.
\end{example}
\vspace{-.1em}
This partitioning assigns each qubit a PZ, which guides all subsequent scheduling. However, the KL partitioning might still result in a two-qubit gate acting on ions whose PZs differ. In such cases, the gate itself is dynamically assigned to the PZ that minimizes the combined estimated shuttling cost for bringing \textit{both} required ions to that specific PZ. This ensures that even \mbox{cross-partition} gates are handled efficiently.
\vspace{-.2em}
\subsection{DAG-based Gate Selection}
\label{subsec:dag_gate_selection}

With partitioning providing a \mbox{high-level} \mbox{qubit-to-PZ} assignment, we next determine the gate execution order, respecting dependencies while promoting parallelism. To do this, we convert the quantum circuit into a \emph{Directed Acyclic Graph} (DAG) using Qiskit~\cite{javadiabhari2024quantum}, where nodes represent gates and edges indicate dependencies. The DAG exposes the \emph{front layer}---gates without preceding dependencies. These gates are mutually commutative and can, in principle, execute in parallel, assuming ion and PZ availability.

Building upon the gate selection strategy from the \mbox{single-zone} setting~\cite[Sec.~IV]{schoenberger2024shuttling}, which selects the single gate with the closest ions, we extend this concept to multiple PZs. For each PZ, we consider all gates in the current DAG front layer that are assigned to it, either because their single qubit was partitioned to that PZ, or because it was determined to be the optimal PZ for a two-qubit gate as described in \autoref{subsec:partitioning}. From this candidate set, we select the \textit{best} gates: the ones whose required ions are currently closest to their PZ.

\begin{example}
   Consider the DAG in \ref{fig:dag}. Given the scenario in~\autoref{ex:partitioning}, we established the partitioned mapping: \mbox{\{PZ1: \{$q_0, q_1$\}, PZ2: \{$q_2$\}\}}. The initial front layer containing nodes \{0: (rz on $q_0$), 6: (rx on $q_1$), 11: (rz on $q_2$)\} is highlighted in~\autoref{fig:dag}. Accordingly, PZ1 considers the gates of nodes 0 and 6. Let the ion representing $q_0$ (for Node 0) be at distance 5 and the ion representing $q_1$ (for Node 6) be at distance 10 from PZ1, then the rz gate of node 0 is selected for PZ1. Since PZ2 considers only the gate of node 11, the rz gate on $q_2$ is selected for PZ2.
\end{example} 

Based on the DAG, a \emph{priority queue} is constructed for each PZ to resolve conflicts and determine which ions to shuttle forward. For a detailed explanation of the priority queue, see~\cite[Sec.~VI]{schoenberger2024shuttling}. This concurrent selection enables the parallel use of all PZs whenever dependencies allow.

\subsection{Orchestration Algorithm}
\label{subsec:orchestration_algorithm}

The combination of partitioning and \mbox{DAG-based} gate selection yields the following algorithm:

\vspace{-1em}
\begin{algorithm}[ht]
\caption{Orchestration Algorithm}
\label{alg:scheduling}
\KwIn{Quantum circuit and initial ion locations}
\KwOut{A schedule $\mathcal{S}$ (shuttling operations, gate executions, and required time steps)}

\textbf{Initialize:}\\
- Partition the ions (mapped to qubits) into sets.\\
- Create the DAG $D=(V,E)$.\\
- Let $\mathcal{S}\gets\emptyset$.\\
- Let $t\gets 0$ (current time step).

\While{$V(D)\neq\emptyset$}{
    \textbf{Identify front layer $F$:} $F=\{\,v\in V(D)\,|\,\nexists\,u\in V(D)\text{ with }(u,v)\in E\}$\;
    
    \ForEach{PZ $p_i$}
        {Identify candidates $C_{p_i}$ from $F$\;
        Select the \emph{best} gate $g_{p_i}$ from $C_{p_i}$\;} 
    Let $G_{\text{done}} \gets \emptyset$ (set of completed gates)\;
    \While{$G_{\text{done}} = \emptyset$}{
        \textbf{Perform Shuttling Step:}\\
        Create shuttling operations (cycles and paths)\;
        Update ion positions\;
        Add shuttling operations to $\mathcal{S}$\;
        \ForEach{PZ $p_i$}{
            \If{ions of $g_{p_i}$ are present in $p_i$}
            {Start gate execution of $g_{p_i}$\;}
            \If{$g_{p_i}$ finished its gate time}{Add $g_{p_i}$ to $G_{\text{done}}$\;}
            }
        $t \gets t + 1$\;
        \textbf{Update DAG:} Remove nodes of gates in $G_{\text{done}}$\;
        }

}
\Return{$\mathcal{S}$}
\end{algorithm}
\vspace{-1em}

This iterative process dynamically adapts the schedule to the evolving ion state and the gates exposed by the DAG. It systematically balances the \mbox{trade-off} between minimizing shuttling through partitioning and maximizing parallelism via concurrent, DAG-aware gate selection, providing a robust compilation strategy for \mbox{multi-zone} QCCD architectures.

\section{Empirical Evaluation}
\label{sec:evaluation}

In this section, we evaluatete the performance of our proposed compilation strategy on \mbox{multi-zone} QCCD architectures. We first evaluate the effectiveness of our \mbox{DAG-based} gate selection technique, demonstrate broad applicability through a comprehensive study of various architectures and different quantum circuits, and finally investigate how multiple PZs impact performance relative to architectures with a single PZ.

\subsection{Experimental Setup}

To evaluate our approach, we use a suite of quantum benchmarks varying in size and structure:

\begin{itemize}
    \item \enquote{GHZ}; prepares the Greenberger–Horne–Zeilinger state,
    \item \enquote{QFT}; scheduling the quantum Fourier transform, and
    \item \enquote{Random}; using a random circuit of up to \mbox{four-qubit} gates which is as deep as wide.
\end{itemize}

MQT Bench~\cite{Quetschlich_2023} circuits are translated via pytket~\cite{Sivarajah_2020} to the native RZZ, RZ, RY, and RX gates used in Quantinuum QCCD devices~\cite{moses2023race}.
Each benchmark is compiled and scheduled using a range of QCCD architectures. The MZ is modeled as a \mbox{grid-type} array connected to one or more linear PZs via \mbox{Y-junctions}. Following the approach of ~\cite{schoenberger2024shuttling}, the grid is described by four values $m,n,v,h$: an $m \times n$ \mbox{grid-graph} with $v$ ($h$) ions between vertical (horizontal) junctions. For example, the graph in \autoref{fig:graph} corresponds to ${4, 4, 3, 3}$.

In our evaluations, up to four PZs are connected to the memory grid. Each PZ can hold up to two ions to allow \mbox{two-qubit} gates, but note that this constraint can be relaxed if future hardware supports larger linear regions. In some research setups, \mbox{two-qubit} gates approach \mbox{single-qubit} speed~\cite{gate_time}, while commercial systems often show a $10$–$100\times$ slowdown. We assume junction traversal takes one time step while shuttling along linear paths is instantaneous. To reflect that gate times are typically faster than junction traversal---without overly penalizing \mbox{two-qubit} operations---we model \mbox{single-qubit} gates as one time step and \mbox{two-qubit} gates as three~\cite{brown2016codesigningscalablequantumcomputer}. These values are configurable in the \mbox{open-source} tool.

All experiments were run on an Intel(R) Xeon(R) W-1370P CPU (@~\SI{3.22}{\giga\hertz}) with \SI{32}{\gibi\byte} RAM using Python 3.8.10. Each benchmark was repeated five times with different seeds to reduce bias from initial ion placement. For all experiments, we initially filled each MZ completely with ions.

\begin{table*}[t]
    \centering
    \caption{Results of the Empirical Evaluation}
    \vspace{-.5em}
    \label{tab:benchmarks}
    \sisetup{
        round-mode=places,
        round-precision=1,
        round-minimum=0.1, %
        table-auto-round, %
        mode=text
    }

    \resizebox{\linewidth}{!}{
        \begin{tabular}{@{}
            c @{} c c %
            @{\hspace{11em}} 
            S[table-format=4.0, parse-numbers=false] %
            S[table-format=4.1] %
            S[table-format=3.1] %
            @{\hspace{6em}} 
            S[table-format=5.0, parse-numbers=false] %
            S[table-format=5.1] %
            S[table-format=5.1] %
            @{\hspace{6em}} 
            S[table-format=5.0, parse-numbers=false] %
            S[table-format=5.1] %
            S[table-format=5.1] %
            @{}}
            \toprule
            \multicolumn{3}{l}{\textbf{Architecture}} &
            \multicolumn{3}{l}{\textbf{GHZ}} &         
            \multicolumn{3}{l}{\textbf{QFT}} &         
            \multicolumn{3}{l}{\textbf{Random}} \\     

            \cmidrule{1-3} \cmidrule{4-6} \cmidrule{7-9} \cmidrule{10-12}

            ~~~~Number of PZs~~~~~ & $m$ $n$ $v$ $h$ & $N$ & %
             \multicolumn{1}{c}{$G$} & \multicolumn{1}{c}{$\hat{T}$} & \multicolumn{1}{l}{$t_\textrm{CPU}$ [s]} &
             \multicolumn{1}{c}{$G$} & \multicolumn{1}{c}{$\hat{T}$} & \multicolumn{1}{l}{$t_\textrm{CPU}$ [s]} &
             \multicolumn{1}{c}{$G$} & \multicolumn{1}{c}{$\hat{T}$} & \multicolumn{1}{l}{$t_\textrm{CPU}$ [s]} \\
            \midrule

        \multirow{8}{*}{\vspace{1.1em}\shortstack[c]{\includegraphics[trim={0 -1.7em 0 1em},clip,angle=0,width=.044\textwidth]{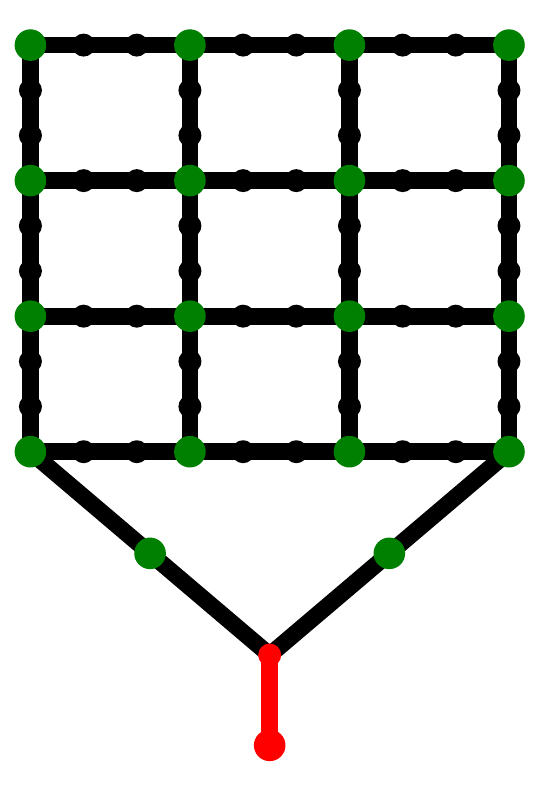} \\Single PZ}}
        & 3 3 1 1 & 12  & 220 & 277.4  & 1.5    & 1039 & 1337.2 & 10.3  & 1340 & 3553.2  & 48.1  \\
        & 4 4 1 1 & 24  & 460 & 603.2  & 6.3    & 4217 & 5476.6 & 139.3  & 6252 & 17601.4 & 1029.4\\ 
        & 3 3 3 3 & 36  & 700 & 890.6  & 16.1   & 10033 & 12639.8& 741.6  & 14914 & 36413.6 & 4466.3\\
        & 5 5 1 1 & 40  & 780 & 1023.2 & 18.0   & 12315 & 15566.2& 1112.1 & 17770 & 45354.8 & 6689.8\\
        & 3 3 5 5 & 60  & 1180 & 1580.2 & 50.8   & 23759 & 30353.2& 4222.1 & 40435 & 52463.7 & 11003.5\\ 
        & 4 4 3 3 & 72  & 1420 & 1854.2 & 74.5   & 30623 & 38622.2& 7149.4 & 59899 & 90722.9 & 15669.0\\ 
        \midrule

        \multirow{8}{*}{\vspace{1.7em}\shortstack[c]{\includegraphics[trim={0 2em 0 1em},clip,angle=0,width=.044\textwidth]{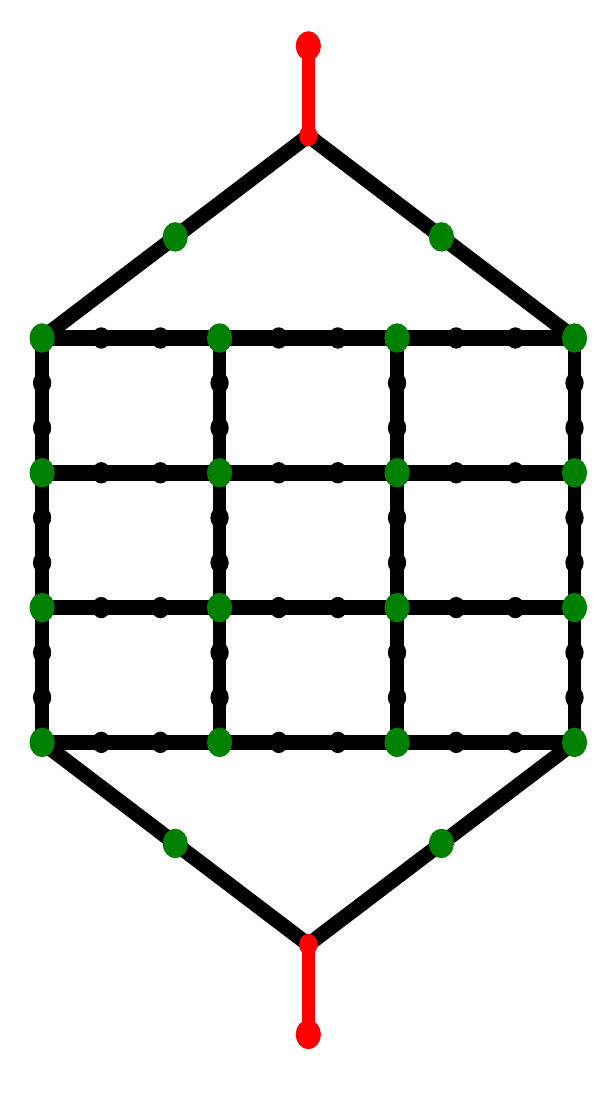}\\Two PZs}}
        & 3 3 1 1 & 12  & 220 & 216.4  & 1.2    & 1039 & 849.4  & 8.3    & 1340 & 2477.6  & 37.2  \\
        & 4 4 1 1 & 24  & 460 & 492.2  & 5.0    & 4217 & 3383.5 & 98.4   & 6252 & 11056.6 & 735.6 \\ 
        & 3 3 3 3 & 36  & 700 & 768.8  & 13.0   & 10033 & 8419.8 & 692.7  & 14914 & 23465.6 & 3248.0\\
        & 5 5 1 1 & 40  & 780 & 819.4  & 13.7   & 12315 & 9641.2 & 894.8  & 17770 & 26733.4 & 4547.1\\ 
        & 3 3 5 5 & 60  & 1180 & 1393.0 & 45.5   & 23759 & 24496.5& 3510.9 & 40435 & 45179.1 & 10956.7\\ 
        & 4 4 3 3 & 72  & 1420 & 1585.8 & 64.4   & 30623 & 25096.1& 5362.7 & 59899 & 61621.1 & 13290.9\\ 
        \midrule

        \multirow{8}{*}{\vspace{2.3em}\shortstack[c]{\includegraphics[trim={2.5em 4em -20em 0},clip,angle=0,width=.082\textwidth]{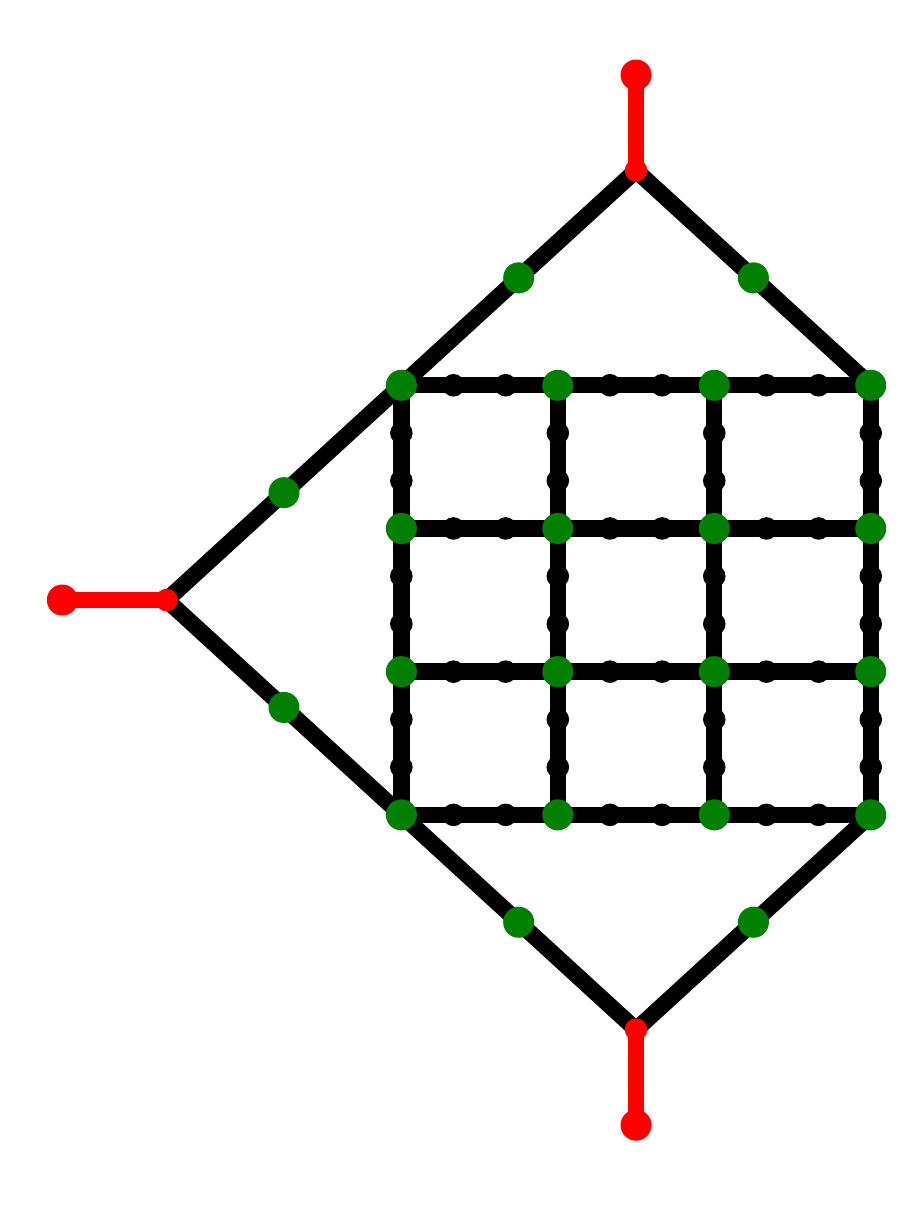}\\Three PZs}}
        & 3 3 1 1 & 12  & 220 & 186.8  & 0.8    & 1039 & 626.8  & 7.1    & 1340 & 2130.0  & 35.5  \\
        & 4 4 1 1 & 24  & 460 & 461.4  & 5.4    & 4217 & 2538.6 & 82.3   & 6252 & 9579.0  & 697.4 \\
        & 3 3 3 3 & 36  & 700 & 722.6  & 13.1   & 10033 & 7052.8 & 507.1  & 14914 & 21296.6 & 3334.1\\ 
        & 5 5 1 1 & 40  & 780 & 810.8  & 13.6   & 12315 & 7059.4 & 662.5  & 17770 & 23341.0 & 4341.3\\ 
        & 3 3 5 5 & 60  & 1180 & 1361.6 & 47.1   & 23759 & 22292.6& 3167.7 & 40435 & 42501.2 & 10672.2\\
        & 4 4 3 3 & 72  & 1420 & 1538.8 & 68.8   & 30623 & 21033.2& 4694.2 & 59899 & 48511.8 & 12043.2\\ 
        \midrule

        \multirow{8}{*}{\vspace{3.0em}\shortstack[c]{\includegraphics[angle=0,width=.083\textwidth]{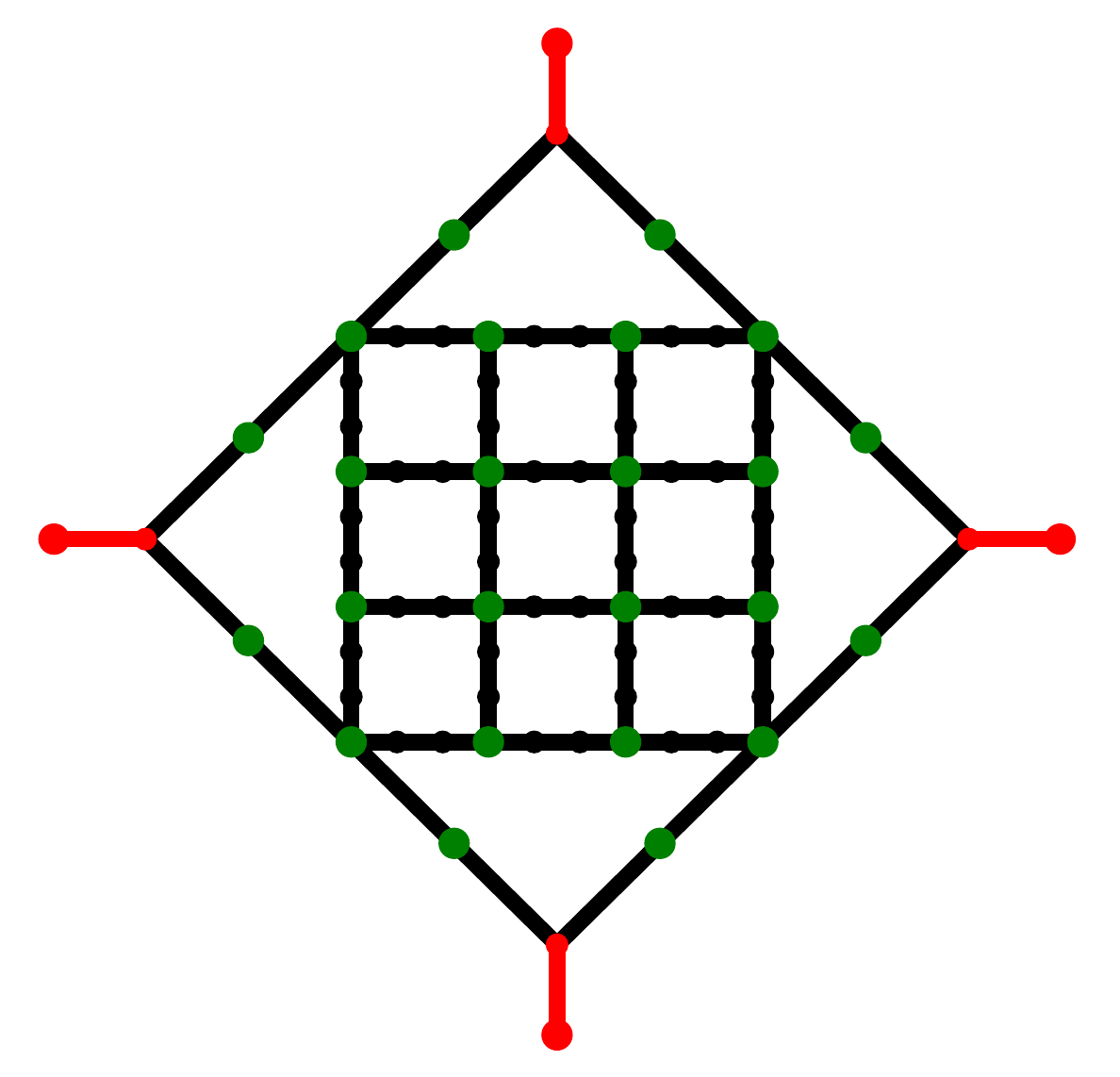}\\Four PZs}}
        & 3 3 1 1 & 12  & 220 & 162.2  & 1.4    & 1039 & 520.6  & 5.1    & 1340 & 1712.5  & 33.9  \\
        & 4 4 1 1 & 24  & 460 & 424.8  & 5.2    & 4217 & 2028.2 & 73.4   & 6252 & 7088.0  & 691.9 \\
        & 3 3 3 3 & 36  & 700 & 708.6  & 13.1   & 10033 & 6514.8 & 487.3  & 14914 & 19355.0 & 3355.3\\ 
        & 5 5 1 1 & 40  & 780 & 738.4  & 14.2   & 12315 & 5842.3 & 596.7  & 17770 & 15803.0 & 4358.1\\
        & 3 3 5 5 & 60  & 1180 & 1360.8 & 52.1   & 23759 & 21705.7& 3828.6 & 40435 & 40630.0 & 10023.9\\
        & 4 4 3 3 & 72  & 1420 & 1520.8 & 73.4   & 30623 & 18591.1& 4369.5 & 59899 & 40822.3 & 10871.0\\

        \bottomrule
        \end{tabular}
    }
    \vspace{-2.4em}
\end{table*}

\subsection{Impact of DAG-Based Gate Selection}

\begin{figure}[t]
    \centering
    \includegraphics[trim={0 0 0 16em},clip,width=.95\linewidth]{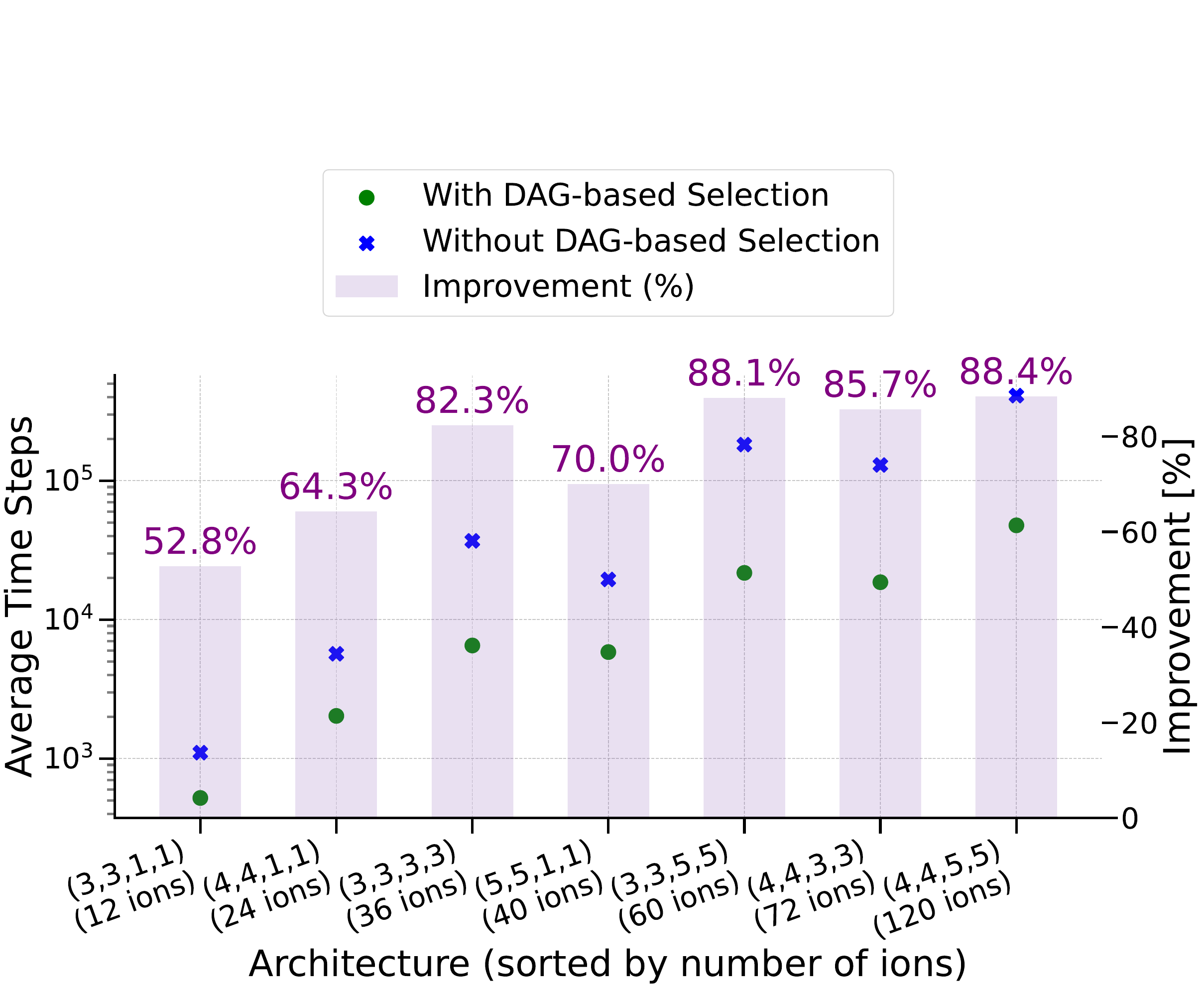}
    \caption{Improvement in time steps executing \enquote{QFT} of using the DAG-based Gate Selection step described in \autoref{subsec:dag_gate_selection}.}
    \label{fig:plot_dag}
    \vspace{-0em}
\end{figure}

\autoref{fig:plot_dag} compares the average execution time (in time steps) achieved with and without the DAG-based gate selection strategy introduced in \autoref{subsec:dag_gate_selection}, evaluated across four representative architectures of varying sizes. For this comparison, we executed the QFT circuit on each architecture on all available qubits. While both evaluations used the qubit partitioning step described in \autoref{subsec:partitioning}, the version without \mbox{DAG-based} selection simply used the input circuit as a fixed sequence of gates and moved through it \mbox{gate-by-gate}.
The scatter plot illustrates the average total time steps required (left vertical axis), while the corresponding bars highlight the percentage improvement (right vertical axis) provided by the \mbox{DAG-based} method. Across all tested benchmarks, the \mbox{DAG-based} scheme consistently reduces the required time steps by between 52\% and 88\%. This demonstrates how \mbox{dependency-aware} selection of upcoming gates, rather than naively moving through the input circuit, can efficiently mitigate unnecessary shuttling.

\subsection{Overall Performance Across QCCD Layouts}

Next, evaluated the broader applicability of our method by compiling our full set of benchmarks (QFT, random, and GHZ) for multiple QCCD layouts, as summarized in \autoref{tab:benchmarks}. Each cell in the table reports the average number of time steps $T$ to complete all gate executions and ion movements for a given circuit with $G$ gates and a specified architecture, along with the CPU time required by our compiler. The architectures are varied in terms of grid size ($m$, $n$, $v$, $h$) and number of PZs, with each architecture holding $N$ ions. The results show that the proposed approach reliably produces valid shuttling schedules across a wide range of QCCD configurations, including both compact and larger \mbox{grid-type} MZs. The implementation is also able to produce efficient shuttling schedules for a single PZ, as well as for multiple PZs.
Furthermore, our implementation is able to efficiently schedule both single- and \mbox{multi-PZ} setups, up to four external PZs. We note that while the tool supports configurations with more than four PZs, we limited this evaluation to a maximum of four to maintain consistency across our comparisons.

\subsection{Impact of Multiple Processing Zones}

\begin{figure}[t]
    \centering
    \includegraphics[trim={0 0 0 -3.5em},clip,width=.95\linewidth]{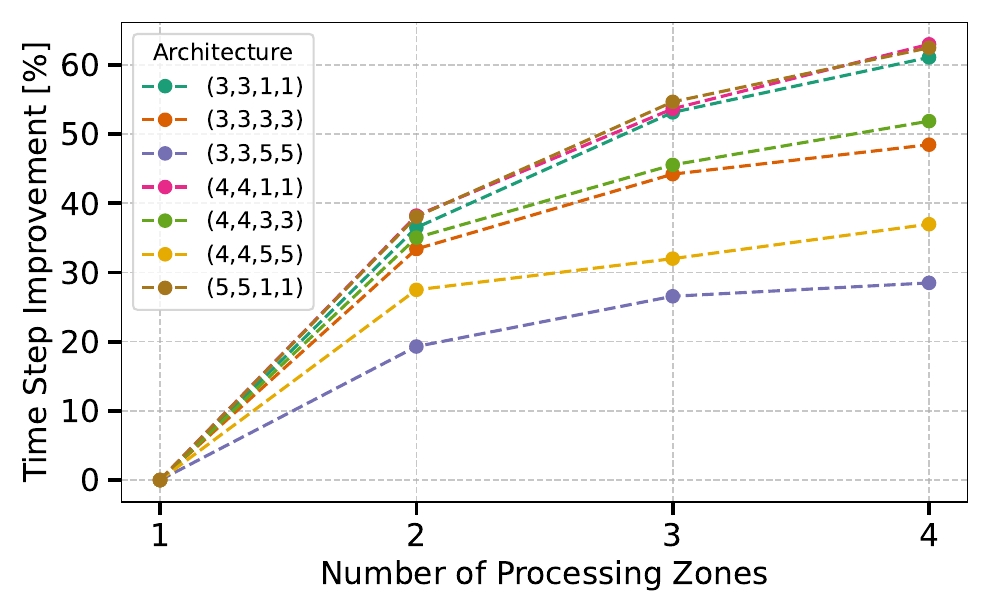}
    \vspace{-.2em}
    \caption{Improvement in time steps executing \enquote{QFT} for an increasing number of PZs.}
    \label{fig:plot_pzs}
    \vspace{-.3em}
\end{figure}

Finally, \autoref{fig:plot_pzs} illustrates how increasing the number of PZs (from one to four) affects the overall schedule length when executing a QFT circuit across several architectures. The \mbox{y-axis} shows the percentage improvement in total execution time relative to the single-zone baseline. As anticipated, adding PZs yields execution time improvements---up to 50–60\%---depending on device geometry and circuit structure. These gains result from parallel gate execution, which helps eliminate wait times and reduce memory grid congestion. However, the benefit of each additional PZ diminishes, indicating reduced returns beyond the initial increase in parallelism.

Collectively, these results demonstrate the applicability and effectiveness of our compilation approach for \mbox{multi-zone} QCCD systems while providing first insights into the benefits of exploiting multiple available PZs in parallel to reduce overall execution runtimes.

\section{Conclusions}
\label{sec:conclusion}

In this work, we presented a comprehensive compilation strategy for QCCD systems with a \mbox{grid-based} MZ and multiple connected PZs. Our approach models PZs as linear regions connected via \mbox{Y-junctions}, moving beyond previous \mbox{black-box} abstractions. We integrated \mbox{path-based} shuttling through PZs into an \mbox{open-source}, \mbox{cycle-based} method for conflict-free transport within the MZ. Central to our method is an orchestration layer that combines strategic qubit partitioning with \mbox{dependency-aware}, concurrent gate selection, enabling efficient operation assignment, reduced \mbox{inter-zone} shuttling, and increased parallelism.
The corresponding tool is available \mbox{open-source} at \url{https://github.com/cda-tum/mqt-ion-shuttler}. Empirical evaluations across three circuits and representative QCCD layouts demonstrated the effectiveness of the proposed approach. The results confirm significant reductions in total execution through the \mbox{dependency-aware} gate selection compared to naively scheduling the input quantum circuit. This allowed us to also assess the impact of scheduling to multiple processing zones.
This work provides a robust and extensible framework for compiling quantum circuits onto \mbox{multi-zone} QCCD architectures, laying a foundation for leveraging future hardware advancements. Future research includes incorporating more realistic noise models, exploring dynamic qubit partitioning, and scheduling \mbox{error-correcting} codes.

\newpage

\section*{\sc Acknowledgments}
This work was supported by the EU's Horizon 2020 programme (DA QC, Grant No. 101001318; MILLENION, Grant No. 101114305), the State of Upper Austria via the COMET programme, the QuantumReady project under Quantum Austria (FFG), and the Munich Quantum Valley, funded by the Bavarian state government through the Hightech Agenda Bayern Plus.

\printbibliography 

\end{document}